\documentclass[%
 reprint,
 amsmath,amssymb,
 aps,
nofootinbib,
showkeys
]{revtex4-2}

\usepackage{graphicx}
\usepackage{dcolumn}
\usepackage{bm}
\usepackage{url,hyperref,lineno,microtype,subcaption}
\usepackage{bm}
\usepackage{braket} 
\usepackage{xcolor} 
\usepackage{makecell} 
\usepackage{float}
\usepackage{ulem} 
\usepackage[utf8]{inputenc}

\newcommand{\SU}[1]{\ensuremath{\mathrm{SU}( #1 )}}

\newcommand{\SO}[1]{\ensuremath{\mathrm{SO}( #1 )}}

\newcommand{\SpR}[1]{\ensuremath{\mathrm{Sp}( #1,\mathbb{R} )}}
\newcommand{\hw}{\ensuremath{\hbar\Omega}}
\newcommand{\ph}[1]{\ensuremath{#1}p-\ensuremath{#1}h}

\newcommand{\Nmax}{\ensuremath{N_{\rm max}}}
\begin{document}

\preprint{APS/123-QED}

\title[SA-EVC Emulators]{\textit{Ab Initio} Symmetry-adapted Emulator \\for Studying Emergent Collectivity \\and Clustering in Nuclei} 

\author{K. S. Becker}
 \email{kbeck13@lsu.edu}
 
\author{K. D. Launey}%
\affiliation{%
 Department of Physics and Astronomy,\\Louisiana State University,\\Baton Rouge, LA, United States
}%

\author{A. Ekström}
\affiliation{%
 Department of Physics,\\Chalmers Institute of Technology,\\Gothenburg, Sweden
}%

\author{T. Dytrych}
\affiliation{%
 Nuclear Physics Institute,\\Czech Academy of Sciences,\\\v{R}e\v{z}, Czech Republic
}%

\begin{abstract}
We discuss emulators from the \textit{ab initio} symmetry-adapted no-core shell-model framework for studying the formation of alpha clustering and collective properties without effective charges. We present a new type of an emulator, one that utilizes the eigenvector continuation  technique but is based on the use of symplectic symmetry considerations. This is achieved by using physically relevant  degrees of freedom, namely, the symmetry-adapted basis, which exploits the almost perfect symplectic symmetry in nuclei. Specifically, we study excitation energies, point-proton root-mean-square radii, along with electric quadrupole moments and transitions for $^6$Li and $^{12}$C. We show that the set of parameterizations of the chiral potential used to train the emulators has no significant effect on predictions of dominant nuclear features, such as shape and the associated symplectic symmetry, along with cluster formation, but slightly varies details that affect collective quadrupole moments, asymptotic normalization coefficients, and alpha partial widths up to a factor of two. This makes these types of emulators important for further constraining the nuclear force for high-precision nuclear structure and reaction observables.
\end{abstract}

\maketitle

\begin{figure*}[th]
  \centering
  \includegraphics[width=0.9\linewidth]{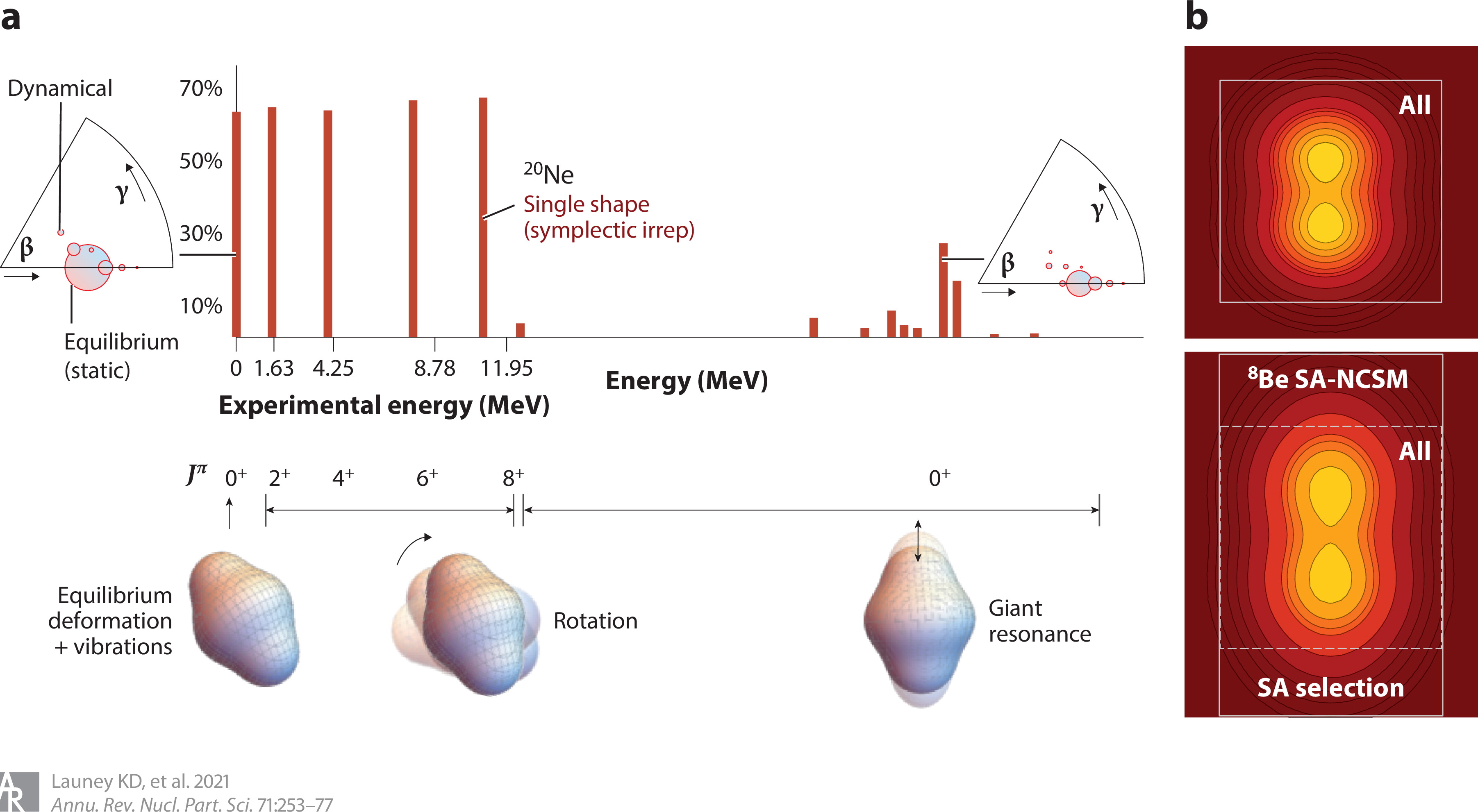}
 \caption{\label{Sp_picture} (a) Emergent symplectic symmetry in nuclei: Contribution of the most dominant shape to the $0^+$ ground state of $^{20}$Ne and its rotational band ($2^+$, $4^+$, $6^+$, and $8^+$), as well as to excited $0^+$ states, pointing to a fragmented giant monopole resonance \cite{DytrychLDRWRBB20}; for selected states, the  deformation distribution within a shape is shown in terms of the shape parameters, the average deformation $\beta$ and triaxiality angle $\gamma$  (based on {\it ab initio} SA-NCSM calculations  with NNLO$_{\rm opt}$  in a model space of 11 HO shells with \hw=15 MeV inter-shell distance). 
(b) Schematic illustration of  the SA concept shown for $^8$Be: a smaller model space (square) includes all possible shapes (labeled as ``All") 
and yields spatially compressed wave functions (top); a larger model space (rectangle in lower panel) accommodates, in a well prescribed way, spatially extended  modes (``SA selection") that are  neglected in smaller model spaces. Figure from Ref. \cite{LauneyMD_ARNPS21}.
}
\end{figure*}

\section{Introduction}
\textit{Ab initio} approaches to nuclear structure and reactions (for an overview, see Ref. \cite{FRIBTAwhite2018}) aim to provide accurate predictions based on few-nucleon forces, such as the ones derived from chiral effective field theory (EFT) [for a review, see e.g. Ref.~\cite{BedaqueVKolck02} and references therein]. To achieve this, it is imperative to utilize high-precision nuclear forces that accurately describe nuclear correlations, from  short- to long-range correlations, as well as to quantify uncertainties that arise from the nuclear force and the controlled approximations in solving the many-body Schr{\"o}dinger equation \cite{Tews_2022}. Such developments employ statistical tools, including, for example, Bayesian analysis \cite{Furnstahl_2015}, global sensitivity methods \cite{Ekstrom19}, and uncertainty estimates based on regression \cite{PhysRevLett.128.202503, PhysRevLett.126.022501}, that sometimes require a large number of computationally intensive calculations which often poses a challenge. 

In this paper, we seek to overcome some of these difficulties by combining the symmetry-adapted no-core shell model (SA-NCSM) framework \cite{DytrychLDRWRBB20,LauneyMD_ARNPS21,LauneyDD16} with the methodology of eigenvector continuation (EVC) \cite{Ekstrom19,PhysRevLett.121.032501,konig20}. The SA-NCSM uses a physically relevant basis that, in manageable model spaces, achieves descriptions of light to medium-mass nuclei, including challenging nuclear features, such as collectivity, clustering, and related continuum effects. Similarly, EVC further reduces the sizes of Hamiltonian matrices by mapping them onto much smaller matrices referred to as emulators, low-dimensional manifolds built upon a set of characteristic solutions to the many-body Schr\"odinger equation. The proposed symmetry-adapted eigenvector continuation (SA-EVC) method opens the door to calculations up through the medium-mass region and studies of collective and clustering nuclear features that otherwise might be computationally infeasible.

With a view toward inferring new knowledge of the nuclear forces relevant to structure and reaction observables, we construct novel SA-EVC emulators to study collective and clustering nuclear properties in $^6$Li and $^{12}$C (an emulator for the $^6$Li binding energy is validated in Ref. \cite{Dj_rv_2022}). Because this study focuses on the method validity, we utilize SA-NCSM calculations for a single harmonic oscillator (HO) strength \hw, for which and for a specific parameterization of the chiral potential we show that the observables under consideration converge with the number of HO excitations, including point-proton root-mean-square (rms) radii and $E2$ transitions. The SA-NCSM utilizes a symplectic \SpR{3}-adapted basis  and selected model spaces\footnote{Throughout the paper, we will refer to the selected SA-NCSM model spaces as SA model spaces.} that are significantly reduced in size due to symmetry considerations without sacrificing the physics of interest. Moreover, we show that the set of chiral potential parameterizations used to train the emulators has no significant effect on dominant nuclear features such as the nuclear shape (and associated symplectic symmetry) and cluster formation, making the SA model spaces highly suitable for this study. However, from one parameterization to another we find that probability amplitudes of wave functions and cluster peak distance vary slightly, affecting by a factor of two or less collective quadrupole moments, asymptotic normalization coefficients (ANCs), and alpha partial widths (which provide the probability for the alpha decay among all possible decays of a state). This suggests that these types of observables, and associated emulators, are important to inform and construct the nuclear forces for high-precision nuclear calculations.

\section{Theoretical Methods}
\subsection{\label{sec:sa-ncsm} \textit{Ab Initio} Symmetry-Adapted\\No-Core Shell Model}

\textit{Ab initio} large-scale calculations \cite{DytrychLDRWRBB20,LauneyMD_ARNPS21} have recently revealed a remarkably ubiquitous and almost perfect symmetry, the \SpR{3} symplectic symmetry, in nuclei that naturally emerges from first principles up through the calcium region (anticipated to hold even stronger in heavy nuclei \cite{Rowe85}). Since this symmetry does not mix nuclear shapes, this novel nuclear feature provides important insight from first principles into the physics of  nuclei and their low-lying excitations as dominated by only one or two collective shapes -- equilibrium shapes with their vibrations -- that rotate (Fig. \ref{Sp_picture}a).  

The SA-NCSM theory \cite{DytrychSBDV_PRL07,LauneyDD16,DytrychLDRWRBB20} capitalizes on these findings and exploits the idea that the infinite Hilbert space can be equivalently spanned by ``microscopic" nuclear shapes and their rotations [or symplectic  irreducible representations (irreps), subspaces that preserve the symmetry], where ``microscopic" refers to the fact that these configurations track with the position and momentum coordinates of each particle. A collective nuclear shape can be viewed as  an equilibrium (``static") deformation and its vibrations (``dynamical" deformations) of the giant-resonance type, as illustrated in the $\beta$-$\gamma$ plots of Fig. \ref{Sp_picture}a~\cite{Rowe13,DytrychLDRWRBB20}. A key ingredient of the SA concept is illustrated in Fig. \ref{Sp_picture}b, namely,  while many  shapes relevant to low-lying states are included in typical shell-model spaces (Fig. \ref{Sp_picture}b, top), the vibrations of largely deformed equilibrium shapes and  spatially extended modes like clustering often lie outside such spaces. The selected model space in the SA-NCSM remedies this, and includes, in a well prescribed way, those  configurations. Note that this is critical for enhanced deformation, since spherical and less deformed shapes, including relevant single-particle effects, easily develop in comparatively small model-space sizes. 

In this study, we utilize the {\it ab initio} SA-NCSM theory \cite{DytrychLDRWRBB20,LauneyMD_ARNPS21,LauneyDD16} that is based on the NCSM concept \cite{NavratilVB00,BarrettNV13} with nuclear  interactions typically derived from the chiral EFT (e.g., \cite{BedaqueVKolck02,EpelbaumNGKMW02,EntemM03,Ekstrom13,Epelbaum:2014sza,PhysRevC.103.054003}). We use SA-NCSM model spaces, which are reorganized to a correlated basis that respects the shape-preserving \SpR{3} symmetry and its embedded symmetry, the deformation-related \SU{3} \cite{DytrychLDRWRBB20,LauneyMD_ARNPS21,LauneyDD16}. We note that while the model utilizes symmetry groups to construct the basis and calculate matrix elements, descriptions are not limited {\it a priori} to any symmetry and can account for significant symmetry breaking.

The SA-NCSM is reviewed in Ref. \cite{LauneyMD_ARNPS21,LauneyDD16} and has been applied to light and medium-mass nuclei using \SU{3}- and \SpR{3}-adapted bases. The  many-nucleon basis states of the SA-NCSM are constructed using efficient group-theoretical algorithms and are labeled according to  \SU{3}$\times$\SU{2} by the proton, neutron and total  intrinsic spins, $S_{\rm p}$, $S_{\rm n}$, and $S$, respectively, and $(\lambda_\omega\,\mu_\omega)$ quantum numbers with $\lambda_\omega=N_z-N_x$ and $\mu_\omega=N_x-N_y$, where $N_x+N_y+N_z=N_0+N$, for a total of $N_0+N$ HO quanta distributed in the $x$, $y$, and $z$ directions\footnote{We follow the notations of Ref. \cite{DytrychSBDV_PRL07}}. Here, $N_0\hw$ is the lowest total HO energy for all particles (``valence-shell configuration") and $N\hw$ ($N\le N_{\rm max}$) is the additional energy of all particle-hole excitations. Thus, for example, $(\lambda_\omega\,\mu_\omega)=(0\, 0)$, for which $N_x=N_y=N_z$, describes a spherical configuration, while $N_z$ larger than  $N_x=N_y$ ($\mu_\omega=0$) indicates prolate deformation. In addition, a closed-shell configuration has $(0\, 0)$. Indeed, spherical shapes, or no deformation, are a part of the SA basis. However, most nuclei --  from light to heavy -- are deformed in the {\it body-fixed} frame, which  for $0^+$ states appear  spherical in the {\it laboratory} frame.

Furthermore, considering the embedding symmetry \SpR{3}$\supset $\SU{3}, one can further organize \SU{3}  deformed configurations into subspaces that preserve \SpR{3} symmetry. Each of these subspaces (symplectic irrep, labeled by $\sigma$) is characterized by  a given equilibrium shape, labeled by a single deformation $N_\sigma(\lambda_\sigma\,\mu_\sigma)$. For example, the symplectic irrep $N_\sigma(\lambda_\sigma\,\mu_\sigma)=0(8\,0)$ in $^{20}$Ne consists of a prolate $0(8\,0)$ equilibrium shape (static deformation) with $\lambda_\omega=8$ and $\mu_\omega=0$ in the valence-shell \ph{0} (0-particle-0-hole) subspace, along with many other \SU{3} deformed configurations or dynamical deformation (vibrations), such as $N_\omega(\lambda_\omega\,\mu_\omega)=2(10\, 0)$, $2(6\,2)$, and $8(16\,0)$, which include particle-hole excitations of the  equilibrium shape to higher shells \cite{DytrychLDRWRBB20,Rowe85,Rowe13}. These vibrations are multiples of 2\hw~\ph{1} excitations of the giant-resonance monopole and quadrupole types, that is, induced by the monopole $r^2=\sum_{i=1}^A{\vec r_i \cdot \vec r_i}$ and quadrupole $Q_{2}=\sqrt{16\pi/5 }\sum_{i=1}^A r_i^2Y_{2}(\hat r_i)$ operators, respectively (for further details, see Refs. \cite{LauneyDD16,LauneyDSBD20}).

An advantage of the SA-NCSM is that  the SA model space can be down-selected from the corresponding ultra-large \Nmax~complete model space to a subset of SA basis states that describe static and dynamical deformation, and within this SA model space the spurious center-of-mass motion can be factored out exactly~\cite{Verhaar60,Hecht71}. Another benefit is the use of group theory for constructing the basis and calculating matrix elements, including the Wigner-Eckart theorem, which allows for calculations with \SU{3} reduced matrix elements that  depend only on $(\lambda\,\mu)$, along with computationally efficacious group-theoretical algorithms and data structures, as detailed in Refs. \cite{DraayerLPL89,LangrDLD18,1937-1632_2019_0_183,DYTRYCH2021108137,MercenneLDEQSD21}. A third advantage is that deformation and collectivity are examined and treated in the approach {\it without} the need for breaking and restoring rotational symmetry. The reason is that basis states utilize the \SU{3}$_{(\lambda\,\mu)}\supset$ \SO{3}$_L$ reduction chain that has a good orbital angular momentum $L$, whereas all \SU{3} reduced matrix elements can be calculated in the simpler canonical \SU{3}$_{(\lambda\,\mu)}\supset$ \SU{2}$_I$  reduction chain (for details, see Ref. \cite{DraayerSU3_1,bahri94rme}). The canonical reduction chain provides a natural reduction to the $x$ and $y$ degrees of freedom, it is simple to work with, and most importantly, provides a complete labeling of a basis state that includes the single-shell quadrupole moment eigenvalue that measures the deformation along the body-fixed symmetry $z$-axis \cite{CARVALHO1986240}. \SU{3} reduced matrix elements calculated within this scheme yield, in turn, matrix elements for the SA-NCSM basis by invoking the Wigner-Eckart theorem with the appropriate \SU{3}$_{(\lambda\,\mu)}\supset$ \SO{3}$_L$ Clebsch-Gordan coefficients that are readily available \cite{DraayerSU3_1}.

We emphasize that all basis states are kept up to some $N_{\rm max}^{\rm C}$, yielding results equivalent to the corresponding $N_{\rm max}^{\rm C}$ NCSM calculations. Building upon this complete $N_{\rm max}^{\rm C}$ model space, we expand the model space to $N_{\rm max}$ by adding selected basis states to include only the necessary vibrations of largely deformed equilibrium shapes that lie outside this $N_{\rm max}^{\rm C}$ (such SA-NCSM model spaces are denoted as $\langle N_{\rm max}^{\rm C}\rangle N_{\rm max} $).

\subsection{Eigenvector Continuation Method in the Symmetry-adapted Framework}

\begin{table*}[th!]
\begin{center}
\begin{tabular}{|c|c|c|c|c|c|c|c|c|c|c|c|c|c|} 
 \hline
& & \multicolumn{6}{c|}{SA} & \multicolumn{6}{c|}{Complete}\\  \hline
 \thead{Nucleus} & \thead{$J^\pi$} & \thead{$N_{\rm max}$} & \thead{Dim} & \thead{$E_{\rm X}$\\$[\text{MeV}]$} & \thead{$r_{\rm rms}$\\$[\text{fm}]$} & \thead{$Q$\\$[e\text{ fm}^2]$} & \thead{$B(E2\uparrow)$\\$[e^2\text{ fm}^4]$} & \thead{$N_{\rm max}$} &  \thead{Dim} & \thead{$E_{\rm X}$\\$[\text{MeV}]$} & \thead{$r_{\rm rms}$\\$[\text{fm}]$} & \thead{$Q$\\$[e\text{ fm}^2]$} & \thead{$B(E2\uparrow)$\\$[e^2\text{ fm}^4]$} \\ [0.5ex] 
 \hline
 \thead{$^{6}$Li} & \thead{$1^+_{\mathrm{g.s.}}$} & \thead{$\langle2_{\text{All}}\rangle 8_{\text{13}}$} &  \thead{$4898$} & \thead{--} & \thead{$2.20$} & \thead{$-0.25$} & \thead{$9.75$} & \thead{$8$} & \thead{$2 \times 10^{5}$} & \thead{--} & \thead{$2.22$} & \thead{$-0.028$} & \thead{$10.04$} \\ 
 \hline
 \thead{$^{6}$Li} & \thead{$3^+_1$} & \thead{$\langle2_{\text{All}}\rangle 8_{13}$} & \thead{$9108$} & \thead{$2.20$} & \thead{$2.20$} & \thead{$-4.12$} & \thead{--} & \thead{$8$} & \thead{$3 \times 10^{5}$} & \thead{$2.65$} & \thead{$2.22$} & \thead{$-4.21$} & \thead{--} \\ 
 \hline
 \thead{$^{12}$C} & \thead{$0^+_{\mathrm{g.s.}}$} & \thead{$ 6_{3}$} & \thead{$552$} & \thead{--} & \thead{$2.41$} & \thead{$0$} & \thead{$35.31$} & \thead{$6$} & \thead{$1 \times 10^{6}$} & \thead{--} & \thead{2.43} & \thead{$0$} & \thead{$35.22$} \\ 
 \hline
 \thead{$^{12}$C} & \thead{$2^+_1$} & \thead{$6_3$} & \thead{$238$} & \thead{$5.73$} & \thead{$2.41$} & \thead{$+5.67$} & \thead{--} & \thead{$6$} & \thead{$5\times 10^6$} & \thead{$3.38$} & \thead{$2.43$} & \thead{$+5.56$} & \thead{--} \\ 
 \hline
\end{tabular}
\caption{\label{tab:model-spaces}
Model space dimensions (labeled as ``Dim"), excitation energy $E_{\textrm X}$, point-proton rms radius $r_{\rm rms}$, electric quadrupole moment $Q$, and $B(E2\uparrow)$ transition strengths from the ground state ($\mathrm{g.s.}$) to the first excited state of $^6$Li and $^{12}$C, calculated with NNLO$_{\rm opt}$ and $\hbar\Omega = 15$ MeV in SA and complete model spaces.  $\langle2_{\text{All}}\rangle 8_{\text{13}}$ denotes an $N_{\text{max}} = 2$ model space with all symplectic irreps (complete), 13 \SpR{3} irreps of which extend to $N_{\text{max}} = 8$; $6_3$ denotes 3 \SpR{3} irreps up to $N_{\text{max}} = 6$.
}
\end{center}
\end{table*}

As introduced in Ref. \cite{PhysRevLett.121.032501}, the EVC method utilizes the fact that if a Hamiltonian is a smooth function of some real-valued parameters, its eigenvectors will also be well-behaved functions of those parameters. In practice, this means that one can use a relatively small number of known wave functions to construct an accurate emulator well-approximated by a low-dimensional manifold, and with it accurately predict observables for an arbitrary chiral potential parameterization \cite{konig20}. To compute these initial wave functions from first principles, it is advantageous to use SA model spaces that can accommodate deformation, including spatially expanded modes, as well as medium-mass regions.

An advantage of the EVC method is that  solutions are achieved by diagonalizing matrices with sizes that are many orders of magnitude smaller than those used in exact calculations. This results in a drastically reduced computational time  with practically no discrepancies from the exact results. EVC thus provides a means of generating large samples of nuclear observables from variations in the Hamiltonian parameters. This, in turn, makes computationally intensive statistical analyses, such as sensitivity studies \cite{Ekstrom19,konig20}, possible. It also allows for a reduced computational load for quantifying uncertainties of \textit{ab initio} predictions.

In this study, we construct emulators capable of probing collective and clustering features by employing the EVC method with SA model spaces.
As illustrated in Table \ref{tab:model-spaces}, the SA-NCSM reduces the sizes of Hamiltonian matrices by up to four orders of magnitude, or equivalently by more than $97\%$. The application of EVC to these SA spaces results in an additional reduction of up to 3 more orders of magnitude, or as much as $99\%$. In this combined framework, the final size of the resulting matrices are as much as $10^{-5}$ times smaller than they would be in the corresponding $N_{\textrm{max}}$ complete spaces. As the first step, we consider a chiral EFT nucleon-nucleon (NN) interaction truncated at next-to-next-to-leading order (NNLO), which depends on 14 low-energy constants (LECs). It turns out that we can write the chiral Hamiltonian as $H(\vec{c}) = \sum_{i=0}^{14} c_i h_i$, where $\vec{c}$ is a vector representing a unique combination of the LECs, $h_i$ are the constituent chiral potentials, $h_0$ is the LEC-independent part of the chiral potential plus relative kinetic energy and the Coulomb interaction, and $c_0 = 1$.

A state $\ket{\psi(\vec{c})}$ can be well-approximated as a linear combination of known ``training'' wave functions $\sum_j^{N_T} \alpha_j (\vec c) \ket{\psi(\vec {c_{T}}_j)}$, where  each $\ket{\psi(\vec {c_{T}}_j)}$ in this study is  the lowest-energy eigenvector of $H(\vec{c_{T}}_j)$ for a given $J^\pi$, $\vec{c_{T}}$ corresponds to a training point in the LEC parameter space, and $N_T$ is the number of training points. The chiral Hamiltonian matrices $h_i$ are constructed in the representation of the training wave functions. These $N_T \times N_T$ matrices are used to emulate the wave function for any set of LECs $\vec{c}$ by solving the Schr\"odinger equation for the unknown $\alpha_j(\vec c)$ as a generalized eigenvalue problem that uses the  norm matrix for the training wave functions, $M_{ij} = \braket{\psi(\vec{c_T}_i)|\psi(\vec{c_T}_j)}$.

The new features here are that we generate the emulator for the electric quadrupole moment $Q$ by constructing the $Q$ matrix in the representation of the training eigenvectors (as done for rms radii in Ref. \cite{Ekstrom19}), and that these are calculated using SA model spaces. The quadrupole moment is then approximated by computing $\braket{\psi(\vec{c})|Q|\psi(\vec{c})}=\sum_{ij}\alpha_i (\vec c) \alpha_j (\vec c) \braket{\psi(\vec{c_T}_i)|Q|\psi(\vec{c_T}_j)}$.

\section{Results and Discussions}

\begin{figure*}[th]
  \centering
  \includegraphics[width=0.48\linewidth]{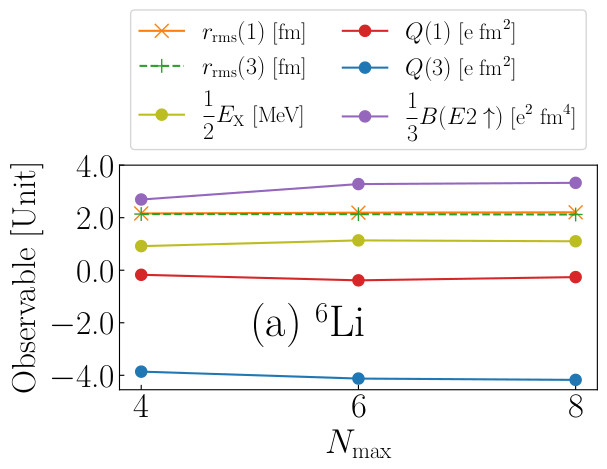}
  \includegraphics[width=0.48\linewidth]{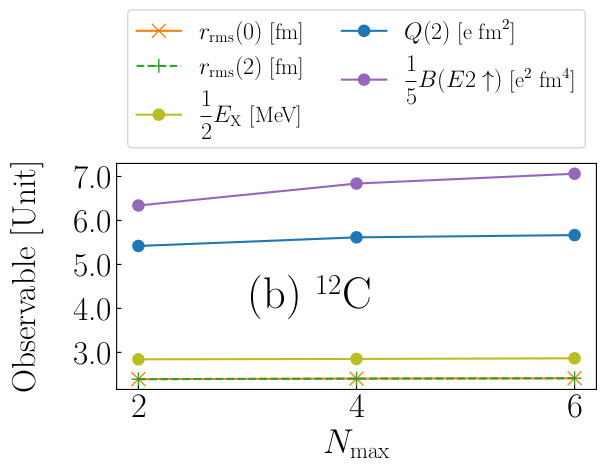}
 \caption{\label{converge_plot} Convergence with $N_{\text{max}}$ of the quadrupole moments $Q(J)$, point-proton rms radii $r_{\rm rms}(J)$, excitation energies $E_{\rm X}$, and $B(E2\uparrow)$ transition strengths for the two lowest-lying states in (a) $^6$Li and (b) $^{12}$C. Observables are computed with the NNLO$_{\text{opt}}$ parameterization for \hw=15 MeV in SA model spaces reported in Table \ref{tab:model-spaces}.
}
\end{figure*}

The results presented in this paper use the SA-NCSM in an \SpR{3} basis with an NN chiral potential up to NNLO as used in \cite{Ekstrom13}. The consistent treatment of NN and three-nucleon (3N) forces at this order is feasible but outside the scope of the present study, which aims to show the validity of the SA-EVC method. We also include the outcomes for a specific NN parameterization, NNLO$_{\rm opt}$ \cite{Ekstrom13}, for which the 3N forces have been shown to contribute minimally to the 3- and 4-nucleon binding energy \cite{Ekstrom13}. Furthermore, the NNLO$_{\rm opt}$ NN potential has been found to reproduce various observables, including the $^4$He electric dipole polarizability \cite{BakerLBND20}; the challenging analyzing power for elastic proton scattering on $^4$He, $^{12}$C, and $^{16}$O \cite{BurrowsEWLMNP19}; neutron-deuteron scattering cross-sections \cite{Miller22}; along with  B(E2) transition strengths  for $^{21}$Mg and $^{21}$F  \cite{Ruotsalainen19} in the SA-NCSM without effective charges. 

For the EVC calculations, we use $N_T=32$ training points within the 14-dimensional parameter space for NNLO. We restrict the ranges of the LECs to lie within $\pm 10\%$ of their  values for NNLO$_{\rm opt}$ \cite{Ekstrom13} and adopt the regularization for NNLO$_{\rm opt}$. We sample training points using a randomly seeded latin hypercube design, and validate the emulators for 256 points that are different from the training points but within the same range of the LECs. 

The SA-EVC results start with SA model spaces that are reduced by three to four orders of magnitude compared to the corresponding $N_{\text{max}}$ complete model space (or, equivalently, NCSM calculations), as outlined in Table \ref{tab:model-spaces}. Moreover, the associated observables are in good agreement for SA and complete model spaces, with differences that are typically comparable to differences resulting from varying \hw~ (see Ref. \cite{DytrychLDRWRBB20}, supplemental material). Specifically, for the example of NNLO$_{\rm opt}$, we report in Table \ref{tab:model-spaces} excitation energies, point-proton rms radii, electric quadrupole moments, and $B(E2\uparrow)$ transition strengths between the two lowest energy states of $^6$Li and $^{12}$C. We also show that for the SA spaces used to train the emulators  all of the above observables are converged with $N_{\text{max}}$ (Fig. \ref{converge_plot}).

\begin{figure*}[th]
    \includegraphics[height=1.62in]{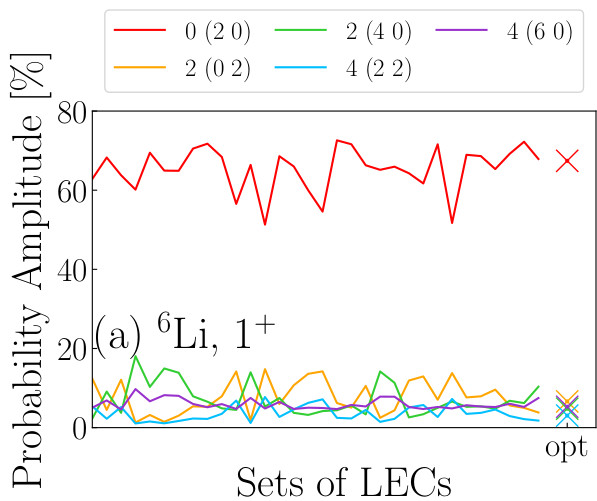}
    \includegraphics[height=1.62in]{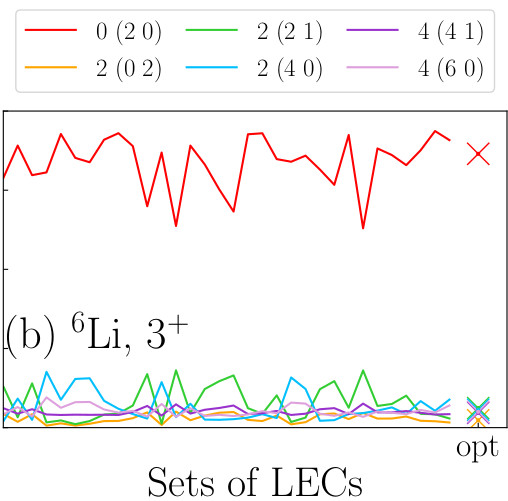}
    \includegraphics[height=1.62in]{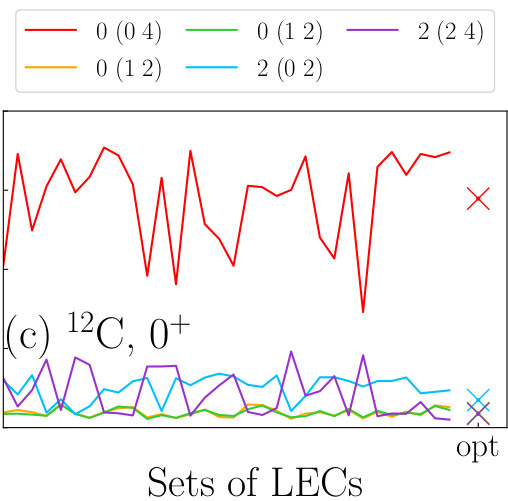}
    \includegraphics[height=1.62in]{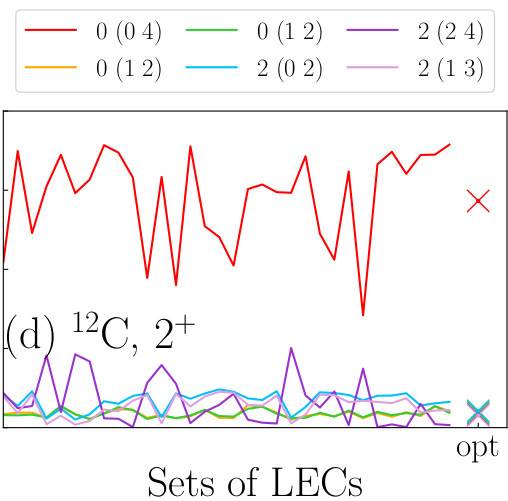}
\caption{\label{wfn_plot} The largest SU(3) probability amplitudes (solid lines) as a function of emulator training LECs sets  for (a) $^6$Li $1^+$ ground state and (b) $^6$Li $3_1^+$ state in $N_{\text{max}}= \langle2_{\rm All}\rangle8_{13}$ model space [all SU(3) states have $\{ S_{\rm p}, S_n, S \} = \{ {1 \over 2},{1 \over 2},1 \}$], as well as for (c) $^{12}$C $0^+$ ground state and (d) $^{12}$C $2_1^+$ state in $N_{\text{max}}=6_3$ [all SU(3) states have $\{ S_{\rm p}, S_n, S \} = \{ 0,0,0 \}$ except for $0 (1\text{ }2)$ with $\{ S_{\rm p}, S_n, S \} = \{ 0,1,1 \}$ (orange) and $\{ 1,0,1 \}$ (green)].
Results are also shown for the NNLO$_{\rm opt}$ parameterization in the corresponding $N_{\rm max}$ complete model space (labeled as ``opt").}
\end{figure*}

Thus, for example, as shown in Table \ref{tab:model-spaces}, collectivity-driven observables agree within 0.3-2.9\%, and radii agree at the sub-percent level. The largest deviation is observed for the $^6$Li $1^+$ quadrupole moment, however, it is important that its sign and very small magnitude are reproduced in both calculations. Furthermore, such differences are expected to decrease in richer model spaces; indeed, in a series of benchmark studies for light nuclei such as $^4$He, $^6$Li, $^{12}$C, and $^{16}$O (reviewed in Ref. \cite{LauneyMD_ARNPS21}), we have shown that the SA-NCSM uses significantly smaller model spaces in comparison to the corresponding large complete \Nmax~model spaces  without compromising the accuracy for various observables (including electron scattering form factors \cite{DytrychHLDMVLO14} and sum rules \cite{BakerLBND20}), as well as for effective inter-cluster potentials \cite{MercenneLDEQSD21}. Ref. \cite{LauneyMD_ARNPS21} has also shown that for light nuclei, the SA-NCSM is in reasonable agreement with other \textit{ab initio} approaches, such as hyperspherical harmonics \cite{Kievsky08,PhysRevLett.89.052502}, NCSM \cite{NavratilVB00,BarrettNV13}, and quantum Monte Carlo \cite{RevModPhys.87.1067}.

\subsection{Collectivity and Clustering of Training Wave Functions}

An important feature of the training wave functions is that the dominant deformed configurations, or the SU(3) content of the states under consideration, remain practically the same for all of the training wave functions (Fig. \ref{wfn_plot}). In addition, the SU(3) content agrees with the probabilities obtained with NNLO$_{\rm opt}$ in the corresponding \Nmax~complete model space, also shown in Fig. \ref{wfn_plot}. This ensures that the same static and dynamical deformed modes  govern the physics for all LECs sets under considerations, thereby justifying the use of the same SA selection for all the training wave functions.

Specifically, we find that one SU(3) irrep dominates  the dynamics of each state at the $50$-$60\%$ level, with several additional configurations each contributing from $1\%$ to $20\%$ depending on the LECs set. Moreover, when the basis states are further organized into \SpR{3} irreps, we find that a single symplectic irrep -- which contains the dominant SU(3) configurations -- contributes at practically the same level from one training wave function to another. For example, the $(2\, 0)$ symplectic irrep in $^6$Li accounts for $83$-$88\%$ of each $1^+$ training wave function, whereas the  $(2\, 0)$ contributes at the $85$-$88\%$ level in the case of the $3^+$, out of thirteen available different irreps. Similarly, the probability of the $(0\, 4)$ irrep in each of the $^{12}$C training ground states is between $80$-$88\%$, and between $82$-$94\%$ for the first $2^+$ states. This is a strong indicator that the emulators are trained on wave functions that retain the symmetry-preserving and symmetry-breaking patterns that are observed in nuclei \cite{DytrychLDRWRBB20} and that the SA model spaces used in this study are sufficient to capture nuclear collectivity. Indeed, the fact that the \SpR{3} symmetry remains a near perfect symmetry for each of the training wave functions, retaining the same shape from one wave function to another, further supports the use of SA selections in the EVC method, or otherwise, the SA model spaces would need to be re-examined.

\begin{figure}[ht]
   \includegraphics[width=\columnwidth]{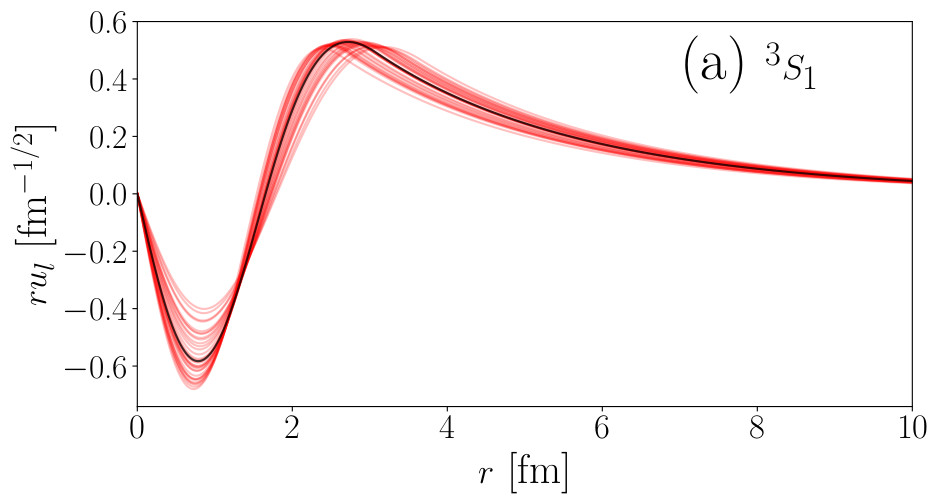}
   \includegraphics[width=\columnwidth]{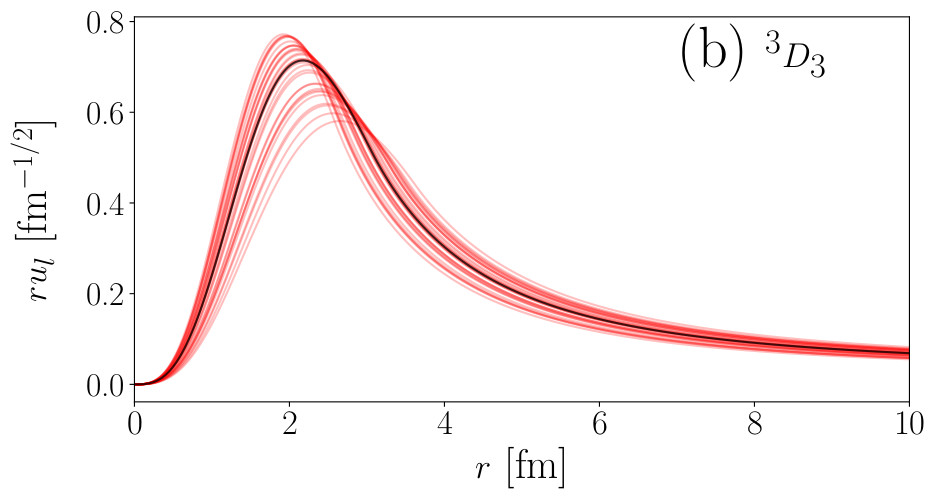}
    \caption{$\alpha + d$ (a) $^3S_1$-wave and (b) $^3D_3$-wave as functions of the relative distance $r$, computed from the $^6$Li training wave functions for SA model spaces reported in Table \ref{tab:model-spaces}. The spread of the curves is given by the $\pm 10\%$ variation in the LECs. The case for NNLO$_{\textrm{opt}}$ is shown in black.}
    \label{fig:6Li_da}
\end{figure}

Another important feature of the training wave functions is that cluster formation is largely unaffected by the choice of interaction parameters. To study this, we project the $^6$Li states onto the $\alpha + d$ system, following Ref. \cite{DreyfussLESBDD20}: we use a ground state for each cluster that is renormalized to the most dominant SU(3) configuration, and we adopt $R$-matrix theory to match the amplitude of the cluster wave function and its derivative to those of the exact Coulomb eigenfunctions at large distances. We note that we are primarily interested in the effect of the LECs on the correlations in the training wave functions; hence, we fix the threshold energy to the experimental one. For the $^3S_1$ partial wave, we observe about $20\%$ variations in the calculated asymptotic normalization coefficients ($C_{0}=1.45$-$2.07$ fm$^{-1/2}$) around their average value and $10\%$ variations in the spectroscopic factor, namely, $SF=0.75$-$0.90$(Fig. \ref{fig:6Li_da}a). This tracks with the $\pm 10\%$ variation in the LECs. For comparison, the NNLO$_{\textrm{opt}}$ ANC for this particular channel is $C_{0}=1.77$ fm$^{-1/2}$ with $SF=0.87$. Interestingly, the height of the second peak, which is located near the nuclear surface and informs the probability of cluster formation, remains fixed for all the parameterizations and coincides with the one for the NNLO$_{\textrm{opt}}$ case, only its position slightly varies with the LECs.

While the $^3D_3$ spectroscopic factors ($SF=0.73$-$0.92$, with $0.90$ for NNLO$_{\rm opt}$)   vary approximately at the $15\%$ level (Fig. \ref{fig:6Li_da}b), which is practically the same as for the $^3S_1$ partial wave, $\alpha$ widths of the $3^+$ state range from $\Gamma_{\alpha}=6.34$ keV to  14.05 keV, which is about $\pm$40\% from  $\Gamma_{\alpha}=9.81$ keV calculated for this particular channel with NNLO$_{\rm opt}$ (similarly to the ANCs, we use the experimental threshold energy). We note that the NNLO$_{\rm opt}$ values for $C_0$ and $\Gamma_{\alpha}$ are reported for a single channel without taking excitations of the clusters into account (e.g., see Ref. \cite{PhysRevLett.114.212502}) and should not be compared directly to experiment. Of particular interest for this study is that the LECs sets induce a change in both the location and magnitude of the peak, to which the probability for alpha decay is typically sensitive to.

To summarize, the behavior of the surface peaks in both channels and the nuclear shapes of the $1^+$ and $3^+$ states in $^6$Li (as well as the shapes of the $0^+$ and $2^+$ states in $^{12}$C) are relatively consistent. This suggests that the terms of the nuclear potential that are independent of the LECs, including parts of the long-range interaction, are largely responsible for cluster formation, along with the development of the nuclear shape [equivalently, almost perfect \SpR{3} symmetry]. In contrast, the LECs, which capture the unresolved short-ranged interactions between nucleons, fine-tune collective and clustering features, and affect the associated observables by only a factor, namely,  1.4 for the $1^+_{\rm g.s.}$ ANCs, 2.2 for the $3^+_1$ alpha width, and 1.4 for the $3^+_1$ quadrupole moment in $^6$Li. Similarly, the quadrupole moment for the $2^+_1$ in $^{12}$C is affected by a factor of 2.1. While the clustering features are explored in this study for the training points only, the SA-EVC approach -- the validation of which is discussed next -- enables uncertainty quantification of such collective and reaction observables if the probability distributions for the LECs are available.

\subsection{Validation of the SA-EVC}

\begin{figure}[th]
  \includegraphics[width=0.78\columnwidth]{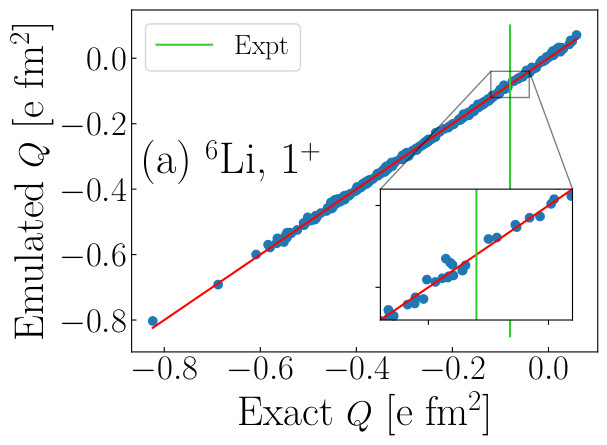}
  \includegraphics[width=0.78\columnwidth]{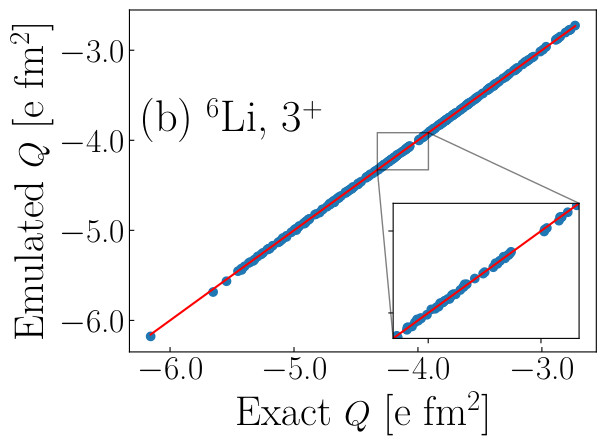}
  \includegraphics[width=0.78\columnwidth]{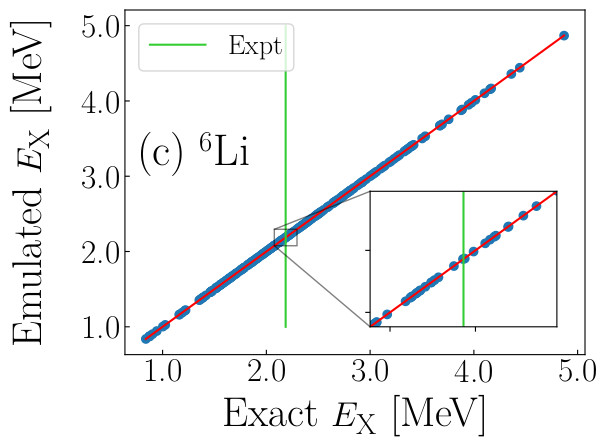}
\caption{\label{6Li_EVC} 
Exact vs. SA-EVC observables in $^{6}$Li  (blue circles) for the quadrupole moment $Q$ of (a) the $1^+$ ground state and (b) the first excited $3^+$ state, as well as (c) for the excitation energy $E_{\rm X}$ of the $3^+$ state, in $\langle2_{\text{All}}\rangle8_{13}$ SA model spaces and for $\hw = 15$ MeV. Also shown is the agreement between the exact and emulated values to guide the eye (red line), and experimental results (vertical green line) where available. Insets show $5\%$-regions surrounding reported experimental data \cite{TILLEY20023} or the NNLO$_{\rm opt}$ result where data is not available [a $50\%$-region is used for the very small $Q$ in (a)].
}
\end{figure}

To validate the SA-EVC approach, we show that for the quadrupole moments of the $^6$Li $1^+$ ground state and first excited $3^+$ state, as well as for the $3^+$ excitation energy, the emulators provide very accurate results compared to the exact outcomes (Fig. \ref{6Li_EVC}). The average relative errors over all $256$ validation LECs sets are respectively $6.91 \times 10^{-2}$, $7.70 \times 10^{-4}$, and $1.20 \times 10^{-4}$. It is clear that any deviations of the emulators from the expected values are negligible, especially considering that, as mentioned above, the SA selection reduces the Hamiltonian dimension by more than $97\%$, and the EVC projection by an additional $99\%$ or more.

It is worth noting that the average error for the ground state quadrupole moment is two orders of magnitude larger than that of the $3^+$ state. We note that $Q(1^+)$ of $^6$Li is very similar in nature to the deuteron quadrupole moment. The extremely small value in both nuclei results from a small mixing of an $L=2$ component into the ground state of $^6$Li  (and of the deuteron), which is not collective in essence like, e.g., the quadrupole moments of the $3^+$ state in $^6$Li or the $2^+$ state in $^{12}$C (discussed below). Indeed, the results of Fig. \ref{6Li_EVC}a reflect the high sensitivity of the underlying NN interaction (and likely 3N forces \cite{FilinMBEKR2021}) to the $L = 2$ mixing in the ground state wave function.

\begin{figure}[th]
  \includegraphics[width=0.78\linewidth]{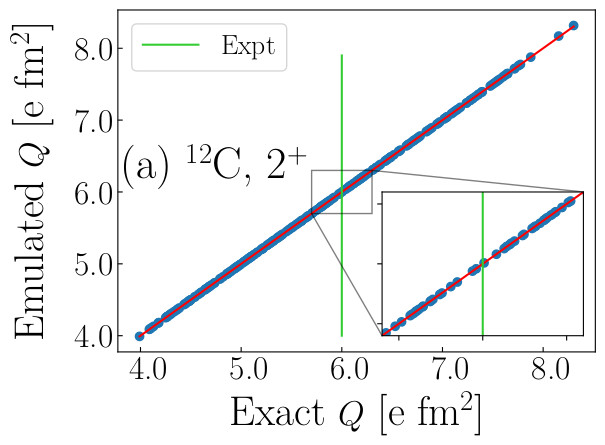}
  \includegraphics[width=0.78\linewidth]{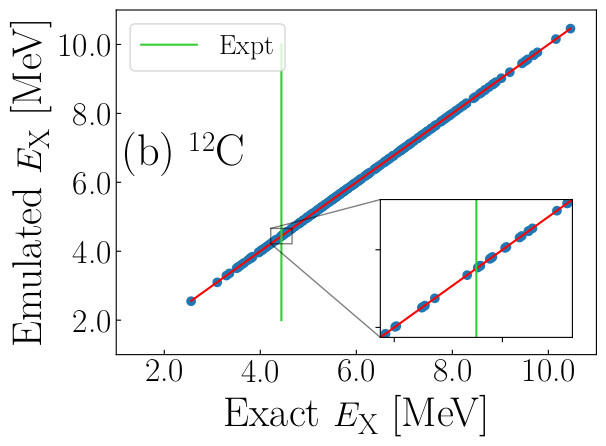}
\caption{\label{12C_EVC}
The same as in Fig. \ref{6Li_EVC} but for (a) the quadrupole moment $Q$ and (b) the excitation energy of the first $2^+$ state in $^{12}$C, calculated in $6_3$ SA model spaces and for $\hw = 15$ MeV. Insets show $5\%$-regions surrounding reported experimental data \cite{KELLEY201771}.}
\end{figure}

Similar to $^6$Li, the SA-EVC emulated $2^+_1$ quadrupole moment and  excitation energy for $^{12}$C are in very close agreement to the exact results (Fig. \ref{12C_EVC}). Namely, the average relative errors are given by $1.02 \times 10^{-4}$ and $6.72 \times 10^{-5}$, respectively. Compared to the average errors reported above for the $3^+_1$ quadrupole moment and excitation energy for $^{6}$Li, we find eight and two times improvement in the emulator's predictions for $^{12}$C, respectively. The reason is likely related to the much smaller SA selection in $^{12}$C and the stronger collective nature observed in the low-lying states of $^{12}$C. Specifically, in $^{6}$Li the SA-EVC uses thousands of basis states, whereas in $^{12}$C  only hundreds of basis states (see Table \ref{tab:model-spaces}). We therefore expect the mixing of configurations to exert a more noticeable effect on $^{6}$Li than on $^{12}$C. The result is that the eigenvectors of $^{12}$C vary in fewer directions than those of $^6$Li, suggesting that more training points for $^6$Li may be beneficial to improve errors. While this warrants further study, this speaks to an advantage of merging the SA and EVC frameworks.

\section{Conclusions}

We have for the first time combined the framework of the SA-NCSM with the EVC procedure into the SA-EVC method for studies of collective and clustering observables. This builds upon earlier SA-NCSM explorations that have shown that an \SpR{3}-adapted model space selection can successfully capture nuclear collectivity while significantly reducing the sizes of Hamiltonian matrices \cite{DytrychLDRWRBB20}. Here, we show that excitation energies, point-proton rms radii, electric quadrupole moments and $E2$ transitions in the two lowest-lying states of $^6$Li and $^{12}$C calculated with the specific parameterization NNLO$_{\rm opt}$ for \hw=15 MeV in SA model spaces are in reasonable agreement with those calculated in the corresponding $N_{\text{max}}$  complete model space (or equally, to NCSM outcomes). We also show that these observables are converged with $N_{\text{max}}$ for the SA selections under consideration. 

Further, we demonstrate that SA-EVC emulators trained on SA model spaces are capable of accurately predicting such observables as the LECs are varied, while further reducing the dimensions of operator matrices by an additional $2$-$3$ orders of magnitude. Combined with the initial reduction provided by the SA-NCSM, the emulator matrices have a dimension as much as $10^{-5}$ times smaller than the corresponding $N_{\textrm{max}}$ complete model spaces. They are small enough to perform linear algebra operations using a single CPU thread on a standard laptop without difficulty. Moreover, the SA-EVC approach will be critical for nuclei beyond the lightest systems; thus, e.g.  in $^{20}$Ne, the complete $\Nmax=8$ model space has dimension of  $1.52\times 10^{11}$, while the {\it ab initio} SA-NCSM  solutions are achieved when using 112 million basis states for $J^\pi=0^+,2^+,4^+$. This can be further reduced to emulators of dimension $10^2$ especially given the predominance of a single symplectic irrep in the ground-state rotational band of this nucleus. Comparing the emulator results to exact calculations performed in the same SA spaces, we find that the average relative errors are typically $10^{-4}$. A larger error ($\sim 10^{-2}$) is found for the quadrupole moment of the $^6$Li ground state, which is highly sensitive to the $L=2$ admixture and hence to the underlying nuclear force, as discussed in the text. A future study that utilizes larger training sets may provide further insight.

In addition to validating the SA-EVC procedure, we show that the symmetry patterns and clustering features in the emulator training wave functions  do not respond strongly to variations in the LECs. Across all of the training wave functions, there is a single nuclear shape (approximate symplectic symmetry) that accounts for $81$-$94\%$ of the total probability. Furthermore, the dominance of important \SU{3} configurations is preserved from one training wave function to another. Projecting the training wave functions for $^6$Li onto the $\alpha + d$ system, we find that the likelihood of cluster formation in both the $^3S_1$- and $^3D_3$-wave channels is largely unaffected by the choice of LECs. Spectroscopic factors, ANCs and $\alpha$-widths extracted from the cluster wave functions all vary within relatively narrow ranges around their average values, ranges that track reasonably well with the $10\%$ variation of the LECs. This suggests that the part of the nuclear potential that is independent of the LECs and is practically the same for all chiral potentials (up to the regularization and related cutoffs employed) provides the dominant features of the wave function, such as \SpR{3} symmetry patterns and clustering formation, while varying the LECs and associated unresolved short-range interactions has an effect on, e.g., collective quadrupole moments, asymptotic normalization coefficients (ANCs), and alpha partial widths up to a factor of two.

In order to better understand the relationships between collectivity and clustering explored in this study, and how both relate to the underlying nuclear forces, sensitivity analyses are required. As we enter the era of high-precision nuclear physics, this is also an important step towards constructing accurate interactions, with quantified uncertainties. We note that properly accounting for clustering features is important for the \textit{ab initio} modeling of nuclear reactions, and related process from fusion to fission. The SA-EVC method provides a clear and now verified framework for generating the huge number of chiral parameterizations required for such analyses. Hence, the door is now open to perform \textit{ab initio} calculations with quantified uncertainties that emerge from the interaction and the controlled many-body approximations, from exotic light nuclei up to medium-mass isotopes, as well from spherical to highly enhanced collective and clustering modes.

\begin{acknowledgments}
We acknowledge invaluable discussions with Jerry P. Draayer, George Rosensteel, David Rowe, and Daniel Langr. This work was supported in part by the U.S. National Science Foundation  (PHY-1913728, PHY-2209060), the European Research Council (ERC) under the European Unions Horizon 2020 research and innovation program (Grant agreement No. 758027), the Czech Science Foundation (22-14497S). KSB greatly appreciates the financial support of a research fellowship from the Louisiana Board of Regents. This work benefited from high performance computational resources provided by LSU (www.hpc.lsu.edu),  the National Energy Research Scientific Computing Center (NERSC), a U.S. Department of Energy Office of Science User Facility operated under Contract No. DE-AC02-05CH11231, as well as the Frontera computing project at the Texas Advanced Computing Center, made possible by National Science Foundation award OAC-1818253.
\end{acknowledgments}

\section*{Data Availability Statement}
The raw data supporting the conclusions of this article will be made available by the authors, without undue reservation.

\section*{Author Contributions}
All authors listed have made a substantial, direct, and intellectual contribution to the work and approved it for publication.

\section*{Conflict of Interest}
The authors declare that the research was conducted in the absence of any commercial or financial relationships that could be construed as a potential conflict of interest.

\nocite{*}

\bibliography{final_bib.bib}

\end{document}